\documentstyle[sprocl,epsf]{article}

\def\be{\begin{equation}}
\def\ee{\end{equation}}
\def\bea{\begin{eqnarray}}
\def\eea{\end{eqnarray}}
\newcommand{\bfx}{\mbox{\bf x}}
\newcommand{\msbar}{\overline{\mbox{\rm MS}}}

\begin{document}

\title{PHASE DIAGRAM OF 3D U(1)+HIGGS THEORY}

\author{ J. PEISA }

\address{Department of Mathematical Sciences, University of Liverpool, \\
Liverpool L69 3BX, UK}




\maketitle\abstracts{We study the properties of the phase transition in
three dimensional U(1)+Higgs theory or Ginzburg-Landau model of
superconductivity. Special attention is paid to large values of scalar
self coupling (Type II superconductors), where the nature of the
transition is unclear. We  present some evidence for an unusual
transition in this regime.} 
  
\section{Introduction}

Three dimensional SU(2)+Higgs theory is an effective theory for
the electroweak sector of the Standard Model at high temperatures and
therefore can be used to describe the electroweak 
phase transition.\cite{intro} It is interesting to compare how the
properties of  system change when one 
replaces the non-abelian gauge group SU(2) with abelian U(1). This
results in U(1)+Higgs theory in three dimensions, which is also known
as Ginzburg-Landau model.

Ginzburg-Landau model is also an effective theory for describing the
superconducting transition between normal and superconducting
states. This gives 
additional motivation for studying the properties of the phase
transition in U(1)+Higgs model.

In this talk I describe numerical simulations I have done in
collaboration with K. Kajantie, M. Laine and
M. Karjalainen.\cite{kklp} 


\section{3d U(1)+Higgs theory}

The three dimensional U(1)+Higgs theory is a locally gauge invariant
continuum U(1) + complex scalar field theory defined by 
\begin{eqnarray*}
S &=& \int d^3x\biggl[\frac{1}{4}(\partial_iA_j-\partial_jA_i)^2+
|(\partial_i+ie_3A_i)\phi|^2 + m_3^2 \phi^*\phi + \lambda_3
\left(\phi^*\phi\right)^2\bigg]. 
\end{eqnarray*}
This theory contains three dimensionful parameters $(e_3, m_3,
\lambda_3)$, so one
can factor out one scale. This leaves only two parameters $
y\equiv m_3^2(e_3^2)/e_3^4,\,\, x\equiv\lambda_3/e_3^2$.

It is well known\,\cite{1storder} that for small values of $x$ (Type I superconductors)
the system has a first order phase transition at some critical point
$y_c(x)$, but the 
situation for large values of $x$ (Type II) is less clear. A schematic phase
diagram is shown in Figure 1, and the main question we are trying to
answer is: how do the properties of the phase transition change when
we move from small values of $x$ to large values of $x$? 

\begin{figure}
\epsfysize=6cm
\centerline{\epsffile{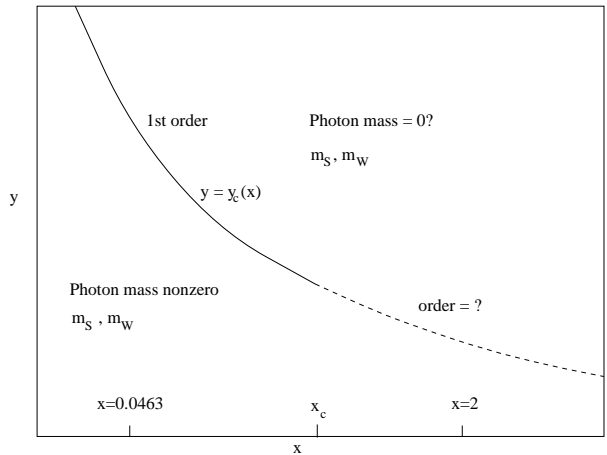}}
\caption{The qualitative phase diagram of the 3d U(1)+Higgs model}
\end{figure}

For large values of $x$ the perturbation theory breaks down and one
has to use non-perturbative methods to study the system. This can be
done by discretizing the system with finite lattice (with lattice
spacing $a$). This results in 
\begin{eqnarray*}
 S &=& \beta_G \sum_{\bfx,i<j} \{ 1- \cos[\alpha_{ij}(\bfx)]\} -\beta_H \sum_{\bfx, i} {\mbox{Re}}\, 
\phi^*(\bfx) U_i(\bfx)\phi(\bfx+\hat{i}) \nonumber \\   
 & & +\sum_{\bfx} \phi^*(\bfx) \phi(\bfx) + \beta_R \sum_{\bfx} \left[\phi^*(\bfx)\phi(\bfx)\right]^2, 
\end{eqnarray*}
where $\beta_G=1/e_3^2a$ and $\alpha_{ij}(\bfx) =
\alpha_i(\bfx)+\alpha_j(\bfx+\hat i)-\alpha_i(\bfx+\hat
j)-\alpha_j(\bfx)$. The relation between other two lattice parameters
can be obtained by relating the 
counter terms in the $\msbar$ and lattice regularisation schemes.\cite{laine}



We focus our studies mainly to the mass spectrum of the system. It
turns out that the mass of the lowest vector state ($m_\gamma$) serves
as an order 
parameter for the system at continuum limit. However, at finite
lattice spacing $a$ this mass cannot be an order parameter. This is
due to the fact that we use compact formulation for 
the gauge field, as it is possible to show\,\cite{polyakov} that then
the monopoles create a mass (which depends on lattice spacing) to this
state. A semiclassical 
calculation can be used to give an estimate for this mass, and we
obtain at $\beta_G = 4$ an extremely small value $m_\gamma/e_3^2=0.003$. 



\section{Simulations}

To study the phase diagram of the system, we have used two values of
$x$: one ($x=0.0463$) corresponding to strongly Type I behaviour and
one ($x=2$) corresponding to strongly Type II. At both points we use
several different lattice sizes and try to work with as small
lattice spacing as possible while still keeping the physical size of
the system reasonably large. 

\begin{figure}
\epsfysize=6cm
\centerline{\epsffile{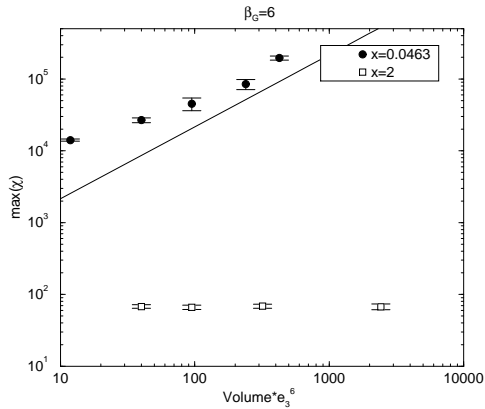}}
\vspace{-0.5cm}
\caption{The maximum of the susceptibility of $\phi^* \phi$ for Type I
and Type II superconductors. The straight line is the expected
behaviour in first order phase transition.}
\end{figure}

As previously mentioned, we concentrate on the masses of the
system. We measure three different operators
\begin{eqnarray*}
O(\bfx)&=&\phi^{\ast}(\bfx)\phi(\bfx), \,\,\, O_i(\bfx)= \mathop{\rm
Im}\phi^{\ast}(\bfx)U_i(\bfx)\phi(\bfx+\hat i), \\
O_{ij}(\bfx)&=&\sin[\alpha_i(\bfx)+\alpha_j(\bfx+\hat i)-
\alpha_i(\bfx+\hat j)-\alpha_j(\bfx)].
\end{eqnarray*}
The masses we extract from asymptotic behaviour of these operators are
labelled $m_H, m_W$ and $m_\gamma$ respectively. To enhance the
projection of these operators to the ground state we 
use blocking. This is absolutely essential in order to get a good signal. 


In addition to the masses, we measure the susceptibility of operator
$\phi^* \phi$. Using finite size scaling, it is possible to
distinguish between first or higher order phase transitions using this
operator. In a first order phase transition, where there are two
different phases, the maximum of susceptibility grows linearly with
volume $\chi  \sim V$. In second order phase transition the expected
behaviour is $\chi \sim V^\alpha$, where $\alpha$ is a critical exponent of the
system. If there is no transition, or the transition is of higher than second
order, or the critical exponent $\alpha \le 0$, the maximum is
constant: $\chi \sim V^0$. 

\subsection{Type I}

We first discuss the transition at small values of $x$. At this regime
we see a very clear two peak structure in $\phi^* \phi$. The
susceptibility calculated from these data clearly shows linear growth
with increasing volume, as can be seen from Figure 2.

Also the mass spectrum of the system shows signal of first
order phase transition. From Figure 3 (left) one can see that all masses have
a clear discontinuity at critical point; furthermore the vector mass
$m_\gamma$ drops to zero at critical point $y_c$, and is consistent with
being zero for $y > y_c$. Thus even though not
rigorously an order parameter, it can be used as an order parameter
for practical purposes.

\begin{figure}
\epsfxsize=6cm
\epsfysize=6cm
\epsffile{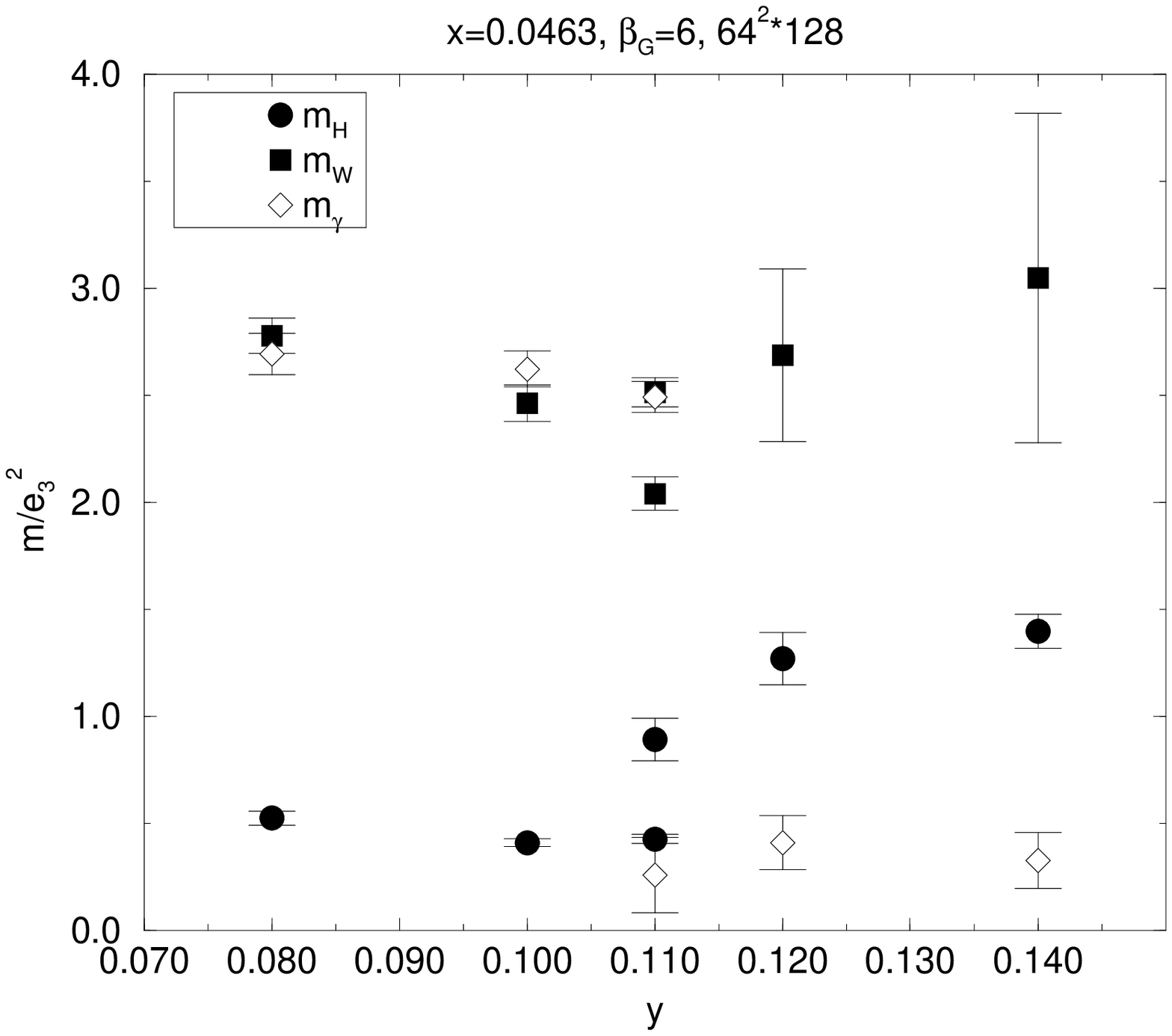}
\vspace{-5.75cm}
\epsfysize=6cm
\epsfxsize=6cm
\hspace{5.75cm}\epsffile{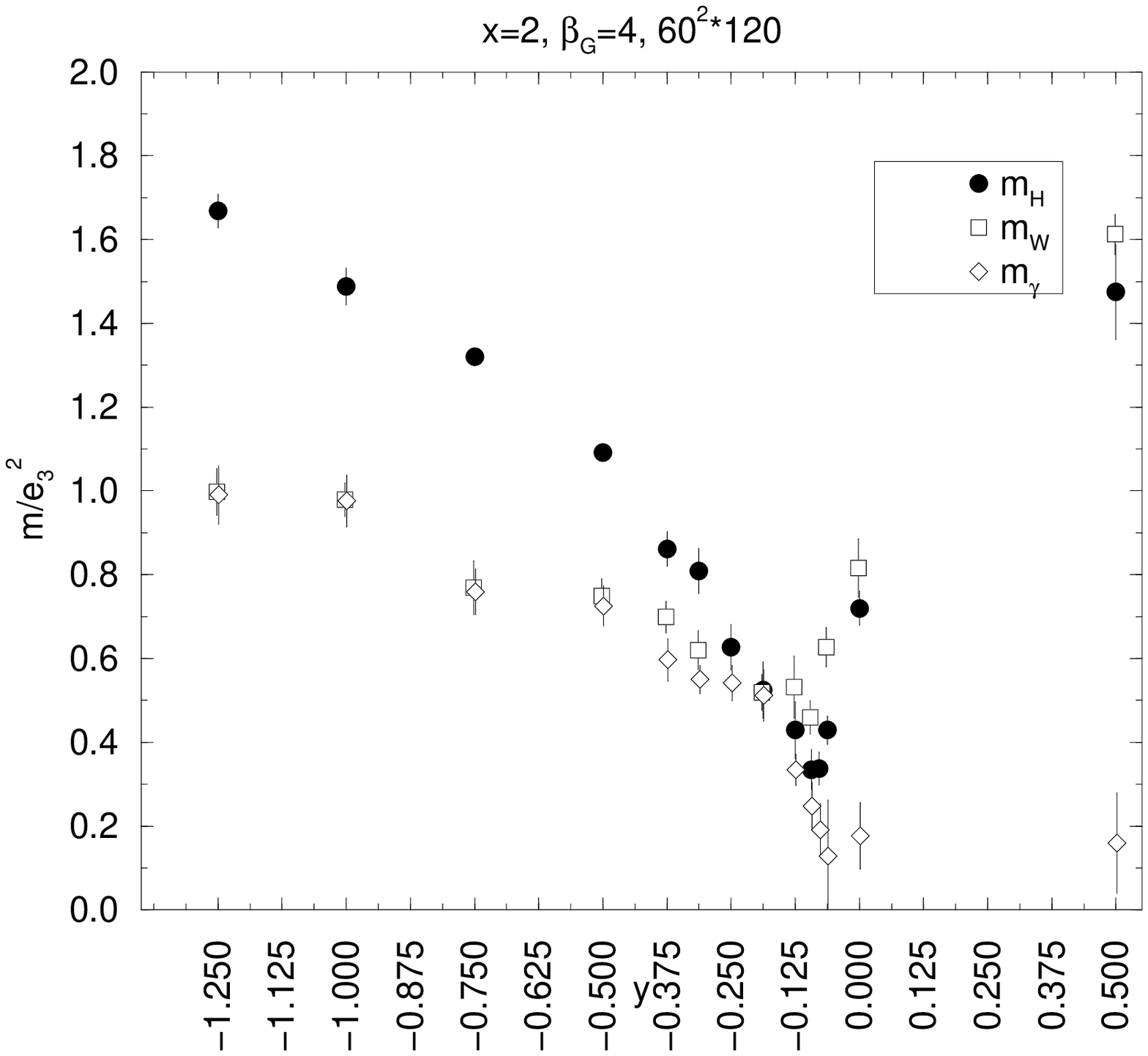}
\vspace{-0.9cm}
\caption{Masses at $x=0.0463$ and at $x=2$.}
\end{figure}

\subsection{Type II}

For large values of $x$ the system behaves in strikingly different
way. It is obvious from Figure 2 that the maximum of the
susceptibility stays constant as one increases volume, which could in
principle mean that there is no phase transition as in
SU(2)+Higgs.\cite{isthere?} However, as previously stated this cannot
be true in 
the continuum limit, so the transition must either be of second order
with critical exponent $\alpha \le 0$ or of higher order. 

If the transition is of second order one would expect that all the
masses vanish at the critical point (as the corresponding correlation
lengths diverge). We have plotted the
three masses we study on Figure 3 (right). One clearly sees that even though
the vector mass $m_\gamma$ goes to zero at critical point the scalar
mass $m_H$ (and the exited state of vector mass $m_W$) remains finite
at $y_c$.  

There are two possible reasons for $m_H$ to stay nonzero even in
second order transition. Firstly
one should check that we are working with such a large volumes that
there are no finite size effects. This is especially important because
there are (nearly) massless exitations in the vicinity of the phase
transition. We have performed simulations with three different lattice
sizes, ranging from $32^3$ to $60^2\times 120$, and see no systematic
evidence for finite size effects.  

The second explanation would be that the masses remain nonzero for any
finite lattice spacing $a$, but vanish at continuum limit. However, we
work with extremely small lattice spacing (near phase transition our
typical correlation length for scalar operator $O$ is roughly ten
times the lattice 
spacing) so one would expect that the results do not change
significantly when taking $a\to 0$. This is further supported by the
fact that the mass of the photon, which is expected to be finite at
nonzero lattice spacing, is withing our statistical errors consistent
with being zero. 


\section{Conclusions}

We have shown evidence towards a non-typical phase transition at Type
II ($x$ large) regime. We find a nonvanishing scalar mass even at
pseudo-critical point even though the vector mass vanishes. Thus the
vector mass serves as an order parameter for the system. The fact that
scalar mass does not vanish means that if the transition is of second
order, it is rather atypical. However, as it is possible that the
vector mass goes to zero continuously, one cannot exclude the
possibility of a second order phase transition. 





\section*{References}
\begin{thebibliography}{99}

\bibitem{intro} M. Laine, these proceedings [hep-ph/9707415].

\bibitem{kklp} K. Kajantie, M. Karjalainen, M. Laine and J. Peisa,
cond-mat/9704056. M. Karjalainen, M. Laine and J. Peisa, Nucl. Phys. B
(Proc. Suppl.) 53 (1997) 475-478. M. Karjalainen and J. Peisa,
hep-lat/9607023.

\bibitem{1storder}
See for example H. Kleinert, {\it Gauge Fields in Condensed Matter},
World Scientific, 1989.

\bibitem{laine} M. Laine, Nucl. Phys. B 451, 484 (1995)

\bibitem{polyakov} A. Polyakov, Phys. Lett. B {\bf 59}, 82 (1975).

\bibitem{isthere?} K. Kajantie, M. Laine, K. Rummukainen and M.
Shaposhnikov,
Phys. Rev. Lett. 77, 2887 (1996) [hep-ph/9605288].

\end{thebibliography}
\end{document}